\documentclass[letterpaper]{article}
\usepackage[preprint]{aaai2027}

\usepackage[hyphens]{url}
\usepackage{graphicx}
\usepackage{natbib}
\usepackage{caption}
\usepackage{amsmath}
\usepackage{amssymb}
\usepackage{algorithm}
\usepackage{algpseudocode}
\usepackage{booktabs}
\usepackage{multirow}
\usepackage{array}
\usepackage{xcolor}
\usepackage{tcolorbox}
\tcbuselibrary{breakable}
\usepackage{tikz}
\usetikzlibrary{arrows.meta, positioning, shapes.geometric, calc}

\definecolor{aggressive}{HTML}{97C1E7}
\definecolor{passive}{HTML}{B66E1A}
\tikzset{baredge/.style={black!45}}

\title{Can Large Language Models Execute Parent Orders?}
\author{
Zane Shen\textsuperscript{\rm 1},
Xinli Xu\textsuperscript{\rm 2},
Guangyi Zhang\textsuperscript{\rm 3},
Jialong Chen\textsuperscript{\rm 4},
Jinsong Zhou\textsuperscript{\rm 2},\\
Cong Chen\textsuperscript{\rm 3},
Guibao Shen\textsuperscript{\rm 2},
Dongyu Yan\textsuperscript{\rm 2},
Luozhou Wang\textsuperscript{\rm 2},
Zhen Yang\textsuperscript{\rm 2}\thanks{Corresponding author.}
}

\affiliations{
\textsuperscript{\rm 1}Independent Researcher,
\textsuperscript{\rm 2}HKUST(GZ),
\textsuperscript{\rm 3}ZJU,
\textsuperscript{\rm 4}SYSU\\
zaneshen1@gmail.com, zheny.cs@gmail.com
}

\begin{document}

\maketitle

\begin{abstract}
Parent-order execution is a core problem in algorithmic trading, where the goal is to split a large order into smaller orders while reducing execution costs. Existing approaches either rely on pre-specified market assumptions that may not hold in practice, or require task-specific training that limits adaptability to new settings. To overcome these limitations, we present the first systematic study of large language models (LLMs) for parent-order execution. This extends the use of LLMs in finance from \textbf{\emph{what}} to trade to \textbf{\emph{how}} to execute. We propose PACE (Plan-Ahead Controlled Execution), a hierarchical framework that decomposes parent-order execution into long-horizon planning and short-horizon execution, requiring neither explicit market assumptions nor task-specific training. Experiments on Shenzhen Stock Exchange Level-1 data show that PACE outperforms TWAP, Almgren-Chriss, and learning-based baselines, exceeding the strongest baseline by \textbf{0.65} bps. Behavioral analysis reveals that LLMs make execution decisions differently from human investors: higher model confidence predicts better performance rather than worse returns, and the model trades earlier rather than procrastinating toward the deadline. These findings suggest that LLMs can complement human traders in execution decisions.
\end{abstract}

\begin{links}
    \link{Code}{https://github.com/zaneopen/PACE}
\end{links}

\section{Introduction}

Reducing execution costs is a central objective for financial institutions~\citep{bertsimas1998optimal}. However, submitting a large order all at once can reveal the trader's intent and leave no room to adjust as the market changes. This may lead to worse traded prices and thus increase execution costs~\citep{kyle1985continuous,o2015high}. Parent-order execution addresses this problem by splitting a large order into smaller orders. The core challenge is to decide how much to trade at each time, aiming to buy lower or sell higher and reduce execution costs.

\begin{figure}[t]
\centering
\includegraphics[width=\linewidth]{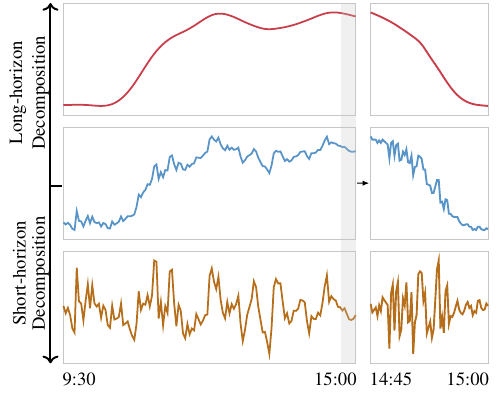}
\caption{Decomposition of a stock price series into its underlying long-horizon trend and residual short-horizon fluctuation using the HP filter~\citep{hodrick1997postwar}.}
\label{fig:price_decomposition}
\end{figure}

Traditional strategies impose simplifying assumptions on market behavior, such as a fixed intraday volume distribution, and trade under these assumptions~\citep{almgren2001optimal,almgren2006bayesian,frei2015optimal}.
Real markets may not follow the assumed form, and model parameters drift over time. Learning-based strategies train data-driven policies but require task-specific rewards, states, and actions~\citep{fang2021universal, lin2021end,wang2021commission,ning2021double,niu2024macmic,xu2025learning}. When market patterns or task specifications change, the policy often needs to be redesigned. An ideal strategy should require neither pre-specified assumptions nor task-specific training, but leverage broad prior knowledge to generate decisions at inference time.

Large language models (LLMs) satisfy these requirements: no pre-specified market assumptions, no task-specific training, and decisions generated at inference time. LLMs have shown strong capabilities in financial text understanding~\citep{huang2023finbert,wu2023bloomberggpt,yang2023fingpt}, factor mining~\citep{wang2026factorminer,han2026quantaalpha}, and trading agents~\citep{xiao2024tradingagents,li2025investorbench,shi2026kronos}. Yet these studies focus on \textbf{\emph{what}} to trade, not \textbf{\emph{how}} to execute. This gap motivates our central question: can LLMs be used for parent-order execution? 

Directly applying LLMs is challenging because price movements are highly noisy. Inspired by the observation illustrated in Fig.~\ref{fig:price_decomposition} that stock prices contain both long-horizon trends and short-horizon fluctuations, we introduce PACE (Plan-Ahead Controlled Execution). It decomposes parent-order execution into two stages: the Planner generates long-horizon plans, while the Executor adjusts trading quantities in response to short-horizon market changes.

We evaluate PACE on Shenzhen Stock Exchange Level-1 data against representative static and learning-based execution strategies, including time-weighted average price (TWAP), Almgren-Chriss (AC), XGBoost, and LSTM baselines, on identical parent orders. The results show that PACE outperforms the strongest baseline by 0.65 basis points (bps), corresponding to USD 6.5 million in annual execution-cost savings for a fund trading USD 100 billion per year.

Behavioral analysis reveals LLM decision patterns strikingly different from human investors: higher Planner confidence predicts better performance, unlike human overconfidence which degrades investment returns~\citep{odean1999investors}, and the Executor places more trades earlier rather than waiting until deadlines, unlike human procrastination~\citep{steel2006integrating}. These differences suggest that LLMs may complement human traders in complex execution decisions.

Our main contributions are as follows:

\begin{itemize}
    \item We provide the first systematic study of LLMs for parent-order execution, extending their financial use from \textbf{\emph{what}} to trade to \textbf{\emph{how}} to execute, with a complete experimental framework for future research and evaluation.

    \item We propose PACE, a framework that separates long-horizon planning from short-horizon execution without pre-specified assumptions or task-specific policy training.

    \item We provide behavioral analysis of LLMs in parent-order execution, linking model actions to confidence and time pressure. This moves beyond performance comparison and helps explain how LLMs behave as execution agents.
\end{itemize}

\section{Related Work}

\textbf{Parent-order execution}.
Existing work on parent-order execution mainly follows static and learning-based lines. Static methods formulate execution through market impact modeling or stochastic control. Representative studies include the mean-variance execution framework of \citet{almgren2001optimal}, the Bayesian adaptive trading model of \citet{almgren2006bayesian}, and the stochastic control formulation by \citet{frei2015optimal}. Later studies improve static strategies through dynamic volume adjustment~\citep{bialkowski2008improving}, or incorporate market microstructure factors~\citep{cartea2015optimal,cartea2016incorporating,tsoukalas2019dynamic}. Learning-based methods instead learn adaptive execution strategies from data. Prior work applies reinforcement learning to execution timing and action selection~\citep{moallemi2022reinforcement,ning2021double,fang2021universal,lin2021end}, uses neural networks to approximate optimal execution solutions~\citep{chen2024parametric}, and explores hierarchical or MoE-based execution strategies~\citep{wang2021commission,niu2024macmic,li2024mixtures,xu2025learning}. These approaches either rely on pre-specified assumptions about price dynamics, trading volume, or market impact, or require task-specific training. In contrast, PACE introduces LLMs to parent-order execution for the first time, combining LLMs' pretrained knowledge with current market observations to generate adaptive decisions at inference time without specifying a price model or training a task-specific policy.

\noindent\textbf{LLMs in finance}. 
Existing LLM research in finance covers three main directions. The first develops financial language models and resources, including FinBERT~\citep{huang2023finbert}, FLANG~\citep{shah2022flue}, BloombergGPT~\citep{wu2023bloomberggpt}, FinGPT~\citep{yang2023fingpt}, and PIXIU~\citep{xie2023pixiu}. The second applies LLMs and foundation models to a broad range of quantitative finance tasks, including factor discovery~\citep{wang2026factorminer,han2026quantaalpha,tang2026alphaagentevo}, financial time-series modeling~\citep{shi2026kronos}, and investment or trading agents~\citep{yu2024fincon,li2024cryptotrade,xiao2024tradingagents,yu2025finmem,ma2025agent}. The third studies evaluation and simulation of financial agents~\citep{li2025investorbench,yang2026twinmarket,tang2026interpreting}. Despite this progress, parent-order execution remains unexplored in LLM-based finance. We fill this gap and provide the first behavioral analysis of LLMs in parent-order execution, revealing several important ways in which their behavior differs from that of human investors.
\section{Method}

\subsection{Preliminary}

In algorithmic trading, parent-order execution is a fundamental problem. The core idea is to split a large order into smaller orders, so that each order consumes less liquidity and reveals less trading intent, thereby reducing execution costs. Parent-order execution operates on a parent order, which specifies the stock ID, trading direction, execution window, and total execution quantity. Given such a parent order, an execution strategy decides when and how much to trade. The resulting time-quantity schedule is called the execution curve. 

% After an execution strategy determines the trading quantity, the gorder-submission style further affects whether and how the order is filled. Aggressive orders seek immediate fills by buying at the Ask1 price or selling at the Bid1 price. Passive orders are posted at the Bid1 price for buys or the Ask1 price for sells; they obtain better prices but may not be filled. Execution performance is often measured in basis points (bps), where 1 bps equals one ten-thousandth. For a fund trading RMB 100 billion annually, a 1 bps execution cost reduction saves about RMB 10 million.

\begin{figure*}[t]
\centering
\includegraphics[width=\textwidth]{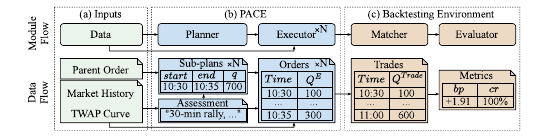}
\caption{Overview of PACE. The lower row gives an example data flow across the three main stages for a parent order that sells 6,000 shares of stock 300308 from 10:30 to 11:00. Here, $N=6$ denotes the number of sub-plans in this example.}
\label{fig:method}
\end{figure*}

\subsection{Inputs}
\label{sec:input}

The inputs consist of three components: the parent order, market history, and the TWAP curve, as shown in Fig.~\ref{fig:method}(a).

\noindent\textbf{Parent order.}
A parent order is specified by $O=(s,d,t_s,t_e,Q)$, where $s$ is the stock ID, $d \in \{\mathrm{BUY},\mathrm{SELL}\}$ is trading direction, $t_s$ and $t_e$ are the start and end times, and $Q$ is the total execution quantity. The parent-order execution window is $[t_s,t_e]$, with duration $T=t_e-t_s$.

\noindent\textbf{Market history.}
The market history contains stock price and traded volume sequences over a recent lookback window. At each time $t$, the market history over the past $L$ minutes:

\begin{equation}
\mathcal{H}_{t-L:t}=\{(P_i,V_i)\}_{i=t-L}^{t}
\end{equation}

where $P_i$ denotes the stock price at minute $i$. $V_i$ denotes the total traded volume at minute $i$. We use the mid-price as a standard proxy for the stock price. At each minute $i$, the mid-price is defined as the average of the Ask1 price $A_i$ and Bid1 price $B_i$. The Ask1 price is the lowest available sell price, and the Bid1 price is the highest available buy price:

\begin{equation}
P_i =
\frac{\text{A}_{i}+\text{B}_{i}}{2}
\end{equation}

\noindent\textbf{TWAP curve.} 
TWAP is a natural baseline which represents the simplest execution curve and provides a reference time-quantity schedule for the execution strategy. For a window $[t_s,t_e]$ with target quantity $Q$ and decision interval $\Delta$, the number of decision times is $K=(t_e-t_s)/\Delta$. TWAP uniformly distributes the total quantity $Q$ across these $K$ decision times. The trading quantity at each decision time is:

\begin{equation}
Q^{\mathrm{TWAP}}=\frac{Q}{K}
\label{eq:qtwap}
\end{equation}

The TWAP curve is therefore defined as the sequence of decision times and corresponding order quantities:

\begin{equation}
\mathcal{C}^{\mathrm{TWAP}}
=
\{(t_k, Q^{\mathrm{TWAP}})\}_{k=1}^{K}
\label{eq:twap_curve}
\end{equation}

\subsection{PACE}
\label{sec:framework}

\noindent\textbf{Motivation.}
Traditional execution strategies are typically either static strategies or learning-based strategies. Static strategies, such as the TWAP curve in Eq.~\ref{eq:twap_curve}, cannot adaptively adjust order quantities in response to market changes. Learning-based strategies lack broader prior knowledge and make order decisions mainly from the observed input data.

To address these limitations, we introduce LLMs into parent-order execution for the first time. In addition, motivated by the observation illustrated in Fig.~\ref{fig:price_decomposition} that financial markets exhibit both coarse price trends and fine local fluctuations, we design PACE to separate long-horizon planning from short-horizon execution. Compared with static strategies, PACE enables timely responses to market changes. Compared with learning-based strategies, PACE leverages LLM prior knowledge beyond the observed inputs.

\noindent\textbf{Long-horizon Planner.}
\label{sec:long_horizon_planner}
Given the parent order $O$, market history $\mathcal{H}_{t_s-T:t_s}$, and TWAP curve $\mathcal{C}^{\mathrm{TWAP}}$, the Planner forms a long-horizon textual trend assessment $\mathcal{R}^{L}$ over the execution window $[t_s,t_e]$, and decomposes the parent order into $N$ execution sub-plans $\{S_n\}_{n=1}^{N}$, as shown in Fig.~\ref{fig:method}(b). Specifically, the Planner partitions the execution window into $N$ shorter time slots of equal duration. For each slot, the LLM outputs a quantity preference score $a_n$ together with an overall confidence score $c$. Here, $a_n \in [-1,1]$, where a larger $a_n$ indicates a preference for allocating more quantity to slot $n$. The confidence score $c \in [0,1]$ reflects the LLM's overall confidence in $\{a_n\}_{n=1}^{N}$. The Planner then maps $\{a_n\}_{n=1}^{N}$ to corresponding quantity-allocation weights as follows:

\begin{equation}
w_n =
(1-\lambda c)
\underbrace{\frac{1}{N}}_{\substack{\text{TWAP}\\\text{allocation}}}
+
\lambda c
\underbrace{
\frac{\exp(a_n)}
{\sum_{j=1}^{N}\exp(a_j)}
}_{\substack{\text{LLM}\\\text{allocation}}}
\label{eq:planner_weight}
\end{equation}

In Eq.~(\ref{eq:planner_weight}), the left underbraced term represents the TWAP allocation, while the right underbraced term represents the LLM allocation. The hyperparameter $\lambda\in[0,1]$ controls the proportion of the LLM allocation. A larger $\lambda$ places more weight on the LLM allocation, whereas a smaller $\lambda$ keeps the allocation closer to TWAP. The Planner then allocates the total quantity $Q$ across time slots according to $w_n$ and produces execution sub-plans $\{S_n\}_{n=1}^{N}$. Each sub-plan $S_n=(u_n,v_n,q_n)$ specifies the start time, end time, and execution quantity of the $n$-th time slot, with duration $\tau_n=v_n-u_n$.

\noindent\textbf{Short-horizon Executor.}
Given the sub-plan $S_n$, the Executor jointly uses market history $\mathcal{H}_{t-\tau:t}$, Planner's long-horizon trend assessment $\mathcal{R}^{L}$, and the TWAP baseline quantity $Q^{\mathrm{TWAP}}$ to accordingly output the order quantity $Q^E_t$:

\begin{equation}
Q^E_t = (1 + {\gamma {z}_t} ) Q^{\mathrm{TWAP}} 
\label{eq:QE}
\end{equation}

Here, $z_t\in[-1,1]$ is the LLM-generated quantity adjustment score, which indicates how the Executor changes the current order quantity relative to the TWAP baseline. A value of $z_t=0$ follows the TWAP baseline, while $z_t>0$ and $z_t<0$ increase and decrease the order quantity, respectively. The hyperparameter $\gamma\in[0,1]$ controls the deviation from TWAP: a smaller $\gamma$ keeps the quantity closer to TWAP, while a larger $\gamma$ amplifies the effect of $z_t$ on the order quantity.

\subsection{Backtesting Environment}
\label{sec:env}

As shown in Fig.~\ref{fig:method}(c), we construct a backtesting environment with two components, the Matcher and the Evaluator, to evaluate the performance of strategies. The Matcher implements the matching mechanism that determines whether strategy-submitted orders are traded, while the Evaluator computes strategy metrics from the traded orders.

\noindent\textbf{Matcher.}
Alg.~\ref{alg:matching} presents the matching mechanism. Each market snapshot provides the Ask1 price $A_t$ and Bid1 price $B_t$ at time $t$. At each snapshot, the strategy $\pi$ submits orders to the Matcher. Each order $o$ has a direction $o.d$, quantity $o.q$, and price $o.p$. The direction $o.d$ follows the parent-order direction, the strategy determines $o.q$ (e.g., PACE uses Eq.~\ref{eq:QE}), and the order-submission setting determines $o.p$.

We consider two order-submission settings. In the aggressive setting, buy orders are submitted at $A_t$ and sell orders are submitted at $B_t$, so they can be traded immediately. In the passive setting, buy orders are submitted at $B_t$ and sell orders are submitted at $A_t$, giving better prices but possibly leaving orders unfilled. The Matcher stores submitted orders in $\mathrm{Orders}$ and checks them against the current $A_t$ and $B_t$. Once an order is traded, the Matcher removes it from $\mathrm{Orders}$.

\begin{algorithm}[t]
\caption{Matching Procedure}
\label{alg:matching}
\begin{algorithmic}
\Require $\mathcal{D}=\{(A_t,B_t)\}$, $\pi$ 
\Statex \quad $\circ$ $\mathcal{D}$: market snapshots
\Statex \quad $\circ$ $A_t,B_t$: Ask1 and Bid1 prices at time $t$
\Statex \quad $\circ$ $\pi$: execution strategy, e.g., TWAP
\State $\mathrm{Orders}\leftarrow[\,]$

\For{each $(A_t,B_t)\in\mathcal{D}$}
    \State $\mathrm{Order}\leftarrow\pi.\mathrm{SubmitOrders}(A_t,B_t)$
    \State $\mathrm{Orders}.\mathrm{extend}(\mathrm{Order})$

    \State {\color{gray}$\triangleright$ orders may remain unfilled}
    \For{each $o\in\mathrm{copy}(\mathrm{Orders})$}
        \State {\color{gray}$\triangleright$ $o.d$ is direction and $o.p$ is price}
        \State $m_B\leftarrow(o.d=\mathrm{BUY}\ \mathrm{and}\ o.p\geq A_t)$
        \State $m_S\leftarrow(o.d=\mathrm{SELL}\ \mathrm{and}\ o.p\leq B_t)$
        \If{$m_B\ \mathrm{or}\ m_S$}
            \State {\color{gray}$\triangleright$ order is traded}
            \State $\Call{Fill}{o}$
            \State $\mathrm{Orders}.\mathrm{remove}(o)$
        \EndIf
    \EndFor
\EndFor
\end{algorithmic}
\end{algorithm}

\noindent\textbf{Evaluator.}
As shown in Fig.~\ref{fig:method}(c), the Evaluator uses two widely used metrics in parent-order execution: price performance ($bp$) and completion rate ($cr$). The $cr$ is defined as:

\begin{equation}
cr=\frac{Q^{\mathrm{Trade}}}{Q}
\label{eq:cr}
\end{equation}

where $Q^{\mathrm{Trade}}$ is the quantity traded by the strategy for the parent order, and $Q$ is the total execution quantity required by the parent order. Thus, a $cr$ of 100\% indicates that the parent order is fully executed. For experiments involving $M$ parent orders, value-weighted price performance $wbp$ is defined as:

\begin{equation}
wbp = \frac{\sum_{i=1}^M P^{\mathrm{TWAP}}_i \cdot Q_i \cdot bp_i}
            {\sum_{j=1}^M P^{\mathrm{TWAP}}_j \cdot Q_j}
\label{eq:wbp}
\end{equation}

A larger $wbp$ indicates better price performance. The unit is basis points (bps), where 1 bps equals one ten-thousandth. Here, $bp$ is the price performance of a single parent order:

\begin{equation}
bp =
\begin{cases}
\dfrac{P^{\mathrm{TWAP}}-P^{\mathrm{Trade}}}
      {P^{\mathrm{TWAP}}}\times 10000,
& d=\mathrm{BUY},\\[6pt]
\dfrac{P^{\mathrm{Trade}}-P^{\mathrm{TWAP}}}
      {P^{\mathrm{TWAP}}}\times 10000,
& d=\mathrm{SELL}
\end{cases}
\label{eq:bp}
\end{equation}

where $P^{\mathrm{Trade}}$ is the strategy's average execution price, $P^{\mathrm{TWAP}}$ is TWAP price over the execution window:

\begin{equation}
P^{\mathrm{TWAP}}=\frac{1}{T}\sum_{i=t_s}^{t_e}P_i
\label{eq:ptwap}
\end{equation}

\section{Experiments}

\subsection{Setup}
\label{sec:setup}

\textbf{Dataset and model selection}. We use Shenzhen Stock Exchange Level-1 Snapshot data covering all trading days from April 2026. For each trading day, we randomly generate 10 parent orders following the procedure detailed in the supplementary material. We experiment with one closed-source model, ChatGPT-5.4, and one open-source model, DeepSeek-v4-flash~\citep{xu2026deepseek}. Both models use their default API parameters. The default reasoning effort is none for ChatGPT-5.4 and high for DeepSeek-v4-flash.

\noindent\textbf{Baselines}. We use TWAP strategy, AC strategy ~\citep{almgren2001optimal}, and learning-based strategies using XGBoost ~\citep{chen2016xgboost} or LSTM ~\citep{hochreiter1997long} as representative baselines. Baseline details are provided in the supplementary material.

\noindent\textbf{Implementation details}. We evaluate two order-submission settings to determine the order price in Alg.~\ref{alg:matching}: all-aggressive and all-passive. In the passive setting, following common industry practice, we cancel unfilled orders and aggressively submit the remaining quantity in the final minute, a procedure known as a \emph{sweep}. To reduce information leakage risk for the LLM, we remove stock IDs and trading dates from the model inputs. Each parent order is repeated eight times to reduce the effect of LLM stochasticity. The main hyperparameters are set to $\lambda=0.3$, $\gamma=0.5$, $\tau=5$ minutes, and $\Delta=1$ minute, where $\lambda$ and $\gamma$ are defined in Eqs.~\ref{eq:planner_weight} and~\ref{eq:QE}, respectively.

% A high completion rate is the primary requirement in parent-order execution. Since passive orders may remain unfilled, all strategies use a $sweep$ operation, a common industry practice, near the end of the parent order: when only the final minute remains, all unfilled orders are canceled, and the remaining quantity is submitted as aggressive orders to improve completion.

\subsection{Main Results}

As shown in Tab.~\ref{tab:main_results}, PACE consistently outperforms all baselines across both settings. Under the aggressive setting, its best variant improves $wbp$ over TWAP by 1.02 bps, and over the strongest baseline by 0.65 bps. Under the passive setting, the corresponding improvements are 1.07 bps and 0.71 bps. Such improvements can be economically meaningful: for a fund trading USD 100 billion annually, replacing the TWAP strategy with PACE can reduce execution costs by about 1 bps, corresponding to approximately USD 10 million in annual savings. In addition, both LLM variants achieve positive gains, suggesting that PACE works across different models.

\begin{table}[t]
\centering
\small
\setlength{\tabcolsep}{3.2pt}
\renewcommand{\arraystretch}{1.28}
\begin{tabular}{cccccc}
\hline
\multirow{2}{*}{Strategy} & \multirow{2}{*}{Method}
& \multicolumn{2}{c}{Aggressive}
& \multicolumn{2}{c}{Passive} \\
\cline{3-6}
& & $wbp$ & gain & $wbp$ & gain \\
\hline
\multirow{2}{*}{Static}
& TWAP & -3.28 & \multicolumn{1}{c}{-} & -4.99 & \multicolumn{1}{c}{-} \\
& AC & -2.93 & +0.35 & -4.74 & +0.25 \\
\hline
\multirow{2}{*}{ML}
& XGB & -3.10 & +0.18 & -4.63 & +0.36 \\
& LSTM & -2.91 & +0.37 & -4.75 & +0.24 \\
\hline
\multirow{2}{*}{Ours}
& GPT-5.4 & -2.76 & +0.52 & -4.49 & +0.50 \\
& DS-v4-f & \textbf{-2.26} & \textbf{+1.02} & \textbf{-3.92} & \textbf{+1.07} \\
\hline
\end{tabular}
\caption{
Main results under aggressive and passive order-submission settings.
Gain denotes the improvement over the TWAP strategy in bps. 
All strategies achieve a 100\% $cr$.
}
\label{tab:main_results}
\end{table}

\subsection{Performance Breakdown}

\noindent\textbf{Significance.}
Using 5,000 bootstrap resamples, DS-v4-f shows significant gains under both aggressive submission (+1.02 bps; 95\% CI: [0.15, 2.12]; $p=0.002$) and passive submission (+1.07 bps; 95\% CI: [0.05, 2.24]; $p=0.014$).

\begin{figure}[t]
\centering
\includegraphics[width=\linewidth]{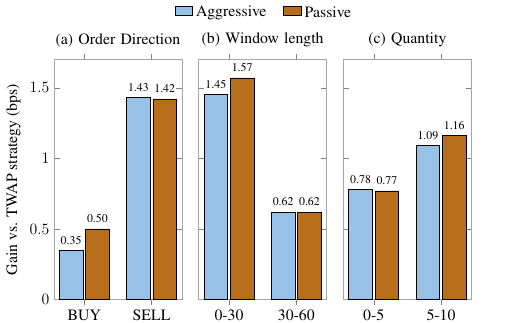}
\caption{Gain of PACE (DS-v4-f) over the TWAP strategy across parent-order characteristics. Window length denotes the execution window length in minutes, and Quantity denotes the parent-order quantity in thousand shares.}
\label{fig:main_group_results}
\end{figure}

\begin{table}[t]
\centering
\small
\setlength{\tabcolsep}{4.6pt}
\renewcommand{\arraystretch}{1.18}
\begin{tabular}{llrr}
\toprule
Type & Method & $wbp$ & gain \\
\midrule
\multirow{3}{*}{Module}
& w/o P+E & -3.28 & \multicolumn{1}{c}{-} \\
& w/o P   & -2.88 & +0.40 \\
& w/o E   & -2.62 & +0.66 \\
\midrule
\multirow{2}{*}{Prompt}
& w/o S & -2.45 & +0.83 \\
& w/o G & -2.38 & +0.90 \\
\midrule
Full & Ours & \textbf{-2.26} & \textbf{+1.02} \\
\bottomrule
\end{tabular}
\caption{
Ablation study in the aggressive setting. P and E denote the Planner and Executor; S and G denote side-specific execution guidance and the market glossary in prompts.
}
\label{tab:module_ablation}
\end{table}

\noindent\textbf{Heterogeneity.}
To further examine whether PACE's gains are consistent across different parent-order characteristics, we split the results along three dimensions: order direction, execution window length, and parent-order quantity. Fig.~\ref{fig:main_group_results} shows that PACE improves over TWAP across all groups. Two patterns are especially notable. First, sell orders show larger gains, which reflect China's short-sale limits: negative information enters prices more slowly, leaving more room for sell execution~\citep{chang2014short}. Second, longer execution windows show smaller gains, because prices become less predictable as the horizon becomes longer.

\noindent\textbf{Volatility Robustness.}
To examine robustness to price noise, we split parent orders into low- and high-volatility groups using stock volatility within the execution window. Higher volatility indicates stronger price fluctuations and thus a noisier execution environment. Fig.~\ref{fig:volatility} shows a clear advantage for PACE under both low and high volatility. This suggests that PACE is more robust to noisy price than ML strategies. Its separation of long-horizon planning from short-horizon execution helps reduce the influence of local noise.

\noindent\textbf{Sweep analysis.}
We examine whether passive-setting gains are driven by the \emph{sweep}. For DS-v4-f, the sweep executes 18.80\% of the quantity, while pre-sweep and sweep trades achieve -3.64 and -6.63 $wbp$, respectively. Thus, the gains arise primarily before rather than from the sweep itself.

\noindent\textbf{Inference cost.}
For the 1,680 parent orders in our main experiment, the total traded value is about USD 35.6 million. For scale, a 1 bps improvement over TWAP corresponds to about USD 3,560 in execution-cost savings. For the DS-v4-f variant, the total LLM API cost is only about USD 30. This suggests that the inference cost of the strongest PACE variant is small relative to its estimated economic benefit.

\subsection{Design Analysis}
\label{sec:ablation}

\noindent\textbf{Module Ablation.}
To analyze the necessity of each module in our method, we conduct ablation study under the all-aggressive setting. w/o P removes long-horizon planning and uniformly allocates the total quantity across sub-plans. w/o E removes short-horizon execution and trades uniformly within each sub-plan. w/o P+E is the TWAP strategy. As shown in Tab.~\ref{tab:module_ablation}, removing either module hurts performance. The Planner contributes more in this setting, but the Executor also adds clear value beyond uniform execution.

% The two module gains are not directly additive, because the Planner changes the sub-plan allocation on which the Executor subsequently operates.

\noindent\textbf{Prompt Ablation.}
To examine whether PACE relies on prompt-specific trading instructions, we remove two prompt components. w/o S removes side-specific guidance, such as ``buy orders prefer lower prices.'' w/o G removes market glossary explanations, such as the meanings of Ask1, Bid1, and TWAP. Both variants still outperform TWAP but underperform the full prompt in Tab.~\ref{tab:module_ablation}. This shows that LLMs retain useful execution ability without these instructions, while clear trading guidance further improves performance.

\begin{figure}[t]
\centering
\includegraphics[width=\linewidth]{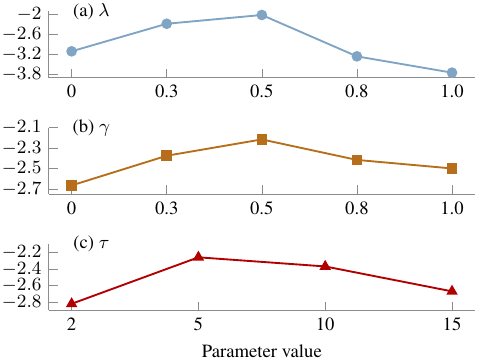}
\caption{Hyperparameter sensitivity of $\lambda$, $\gamma$, and $\tau$ under the aggressive setting, where the y-axis reports $wbp$ in bps.}
\label{fig:ablation}
\end{figure}

\begin{figure}[t]
\centering
\includegraphics[width=\linewidth]{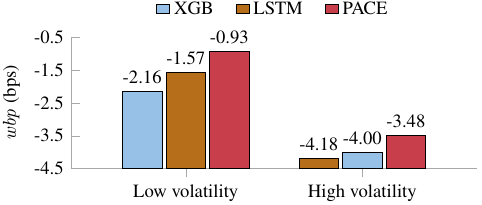}
\caption{$wbp$ in low- and high-volatility scenarios.}
\label{fig:volatility}
\end{figure}

\noindent\textbf{Hyperparameter Sensitivity.}
We examine the sensitivity of PACE to $\lambda$ in Eq.~\ref{eq:planner_weight}, $\gamma$ in Eq.~\ref{eq:QE}, and the sub-plan duration $\tau$ under the aggressive setting. Fig.~\ref{fig:ablation} shows that PACE performs best at moderate parameter values, while extreme values reduce overall performance. The parameters $\lambda$ and $\gamma$ directly control the strength of LLM signals. If they are too small, PACE stays close to TWAP; if they are too large, PACE may overreact to noisy signals. A very small $\tau$ creates too many sub-plans, making Planner allocations too fine-grained and noise-sensitive. A very large $\tau$ creates overly long sub-plans, making Executor adjustments more exposed to noise.

\noindent\textbf{Planner replanning.}
By default, the Planner is called only once when the parent order starts. We test whether updating the Planner during execution can improve performance. In the replanning variant, the Planner dynamically uses the latest market history to generate a new plan for the remaining execution window after each sub-plan ends. Tab.~\ref{tab:replanning_test} shows that replanning hurts shorter parent orders but improves longer parent orders. For shorter parent orders, frequent replanning may disrupt the trading pace and introduce extra noise. For longer parent orders, one-shot planning becomes harder because future trends are less predictable, while replanning can dynamically adjust later sub-plans to new market changes.

\begin{table}[t]
\centering
\small
\setlength{\tabcolsep}{7pt}
\begin{tabular}{ccc}
\toprule
Window length & One-shot & Replanning \\
\midrule
10--40 min & \textbf{-1.99} & -2.50 \\
50--60 min & -2.78 & \textbf{-2.48} \\
\bottomrule
\end{tabular}
\caption{Planner replanning test. Values are reported in $wbp$.}
\label{tab:replanning_test}
\end{table}

\begin{table}[t]
\centering
\small
\setlength{\tabcolsep}{2.4pt}
\renewcommand{\arraystretch}{1.28}
\begin{tabular}{c*{6}{c}}
\toprule
& \multicolumn{2}{c}{GPT-5.4 none}
& \multicolumn{2}{c}{DS-v4-f none}
& \multicolumn{2}{c}{DS-v4-f high} \\
\midrule
$c$
& 6.96$^{**}$
& 7.34$^{***}$
& 5.03$^{**}$
& 5.68$^{**}$
& 4.78$^{**}$
& 5.42$^{**}$ \\
{}
& (2.39)
& (2.58)
& (2.14)
& (2.39)
& (2.04)
& (2.22) \\
Controls
& NO & YES
& NO & YES
& NO & YES \\
\midrule
Adj. $R^2$
& 0.020 & 0.041
& 0.022 & 0.025
& 0.020 & 0.021 \\
Obs.
& 1680 & 1680
& 1680 & 1680
& 1680 & 1680 \\
\bottomrule
\end{tabular}
\caption{Planner confidence and execution performance. The dependent variable is $bp$. None/high denote reasoning-effort. Parentheses report $t$-statistics based on parent-order-clustered standard errors. $^{***}p<0.01$ and $^{**}p<0.05$.}
\label{tab:confidence_performance_regression}
\end{table}

\subsection{Case Study}

\begin{figure*}[t]
\centering
\includegraphics[width=\textwidth]{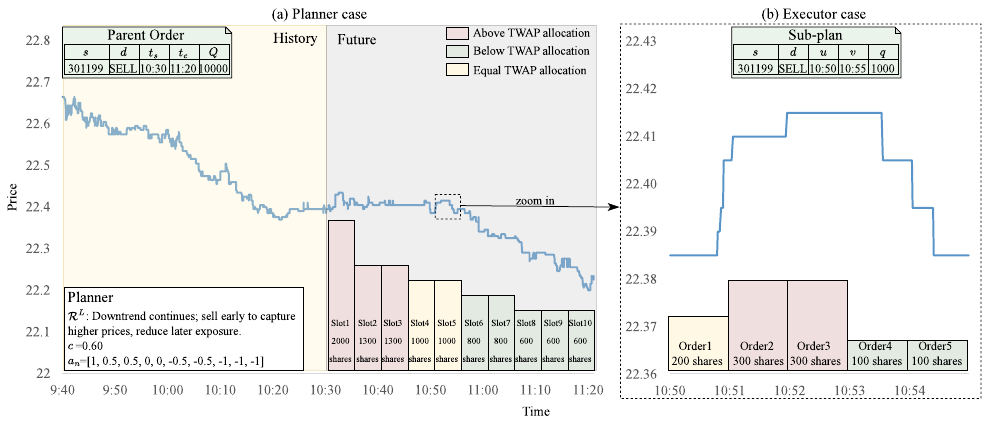}
\caption{Case study of PACE. (a) Using only pre-execution history, the Planner front-loads the sell order before the realized price decline. (b) Using only information available at each decision time, the Executor trades more at locally higher prices.}
\label{fig:case}
\end{figure*}

Fig.~\ref{fig:case}(a) presents a sell parent order from 10:30 to 11:20. Based on the market history from 9:40 to 10:30, the Planner expects the downward trend to continue and allocates more quantity to the early slots. The realized price keeps declining after the parent order starts, so this front-loaded allocation helps the sell order trade at higher prices. Fig.~\ref{fig:case}(b) presents a sell sub-plan from 10:50 to 10:55. The Executor increases the sell quantity from the TWAP baseline of 200 shares to 300 shares when prices are relatively high, and reduces it to 100 shares when prices decline. As a result, it sells more shares at more favorable prices and lowers execution costs.

\subsection{Planner Behavior}

\begin{figure}[t]
\centering
\includegraphics[width=\linewidth]{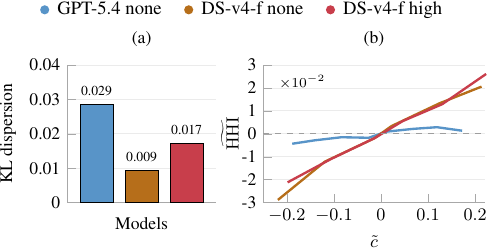}
\caption{Planner behavior. (a) Lower KL dispersion means more stable long-horizon scores. (b) $\tilde{c}$ and $\widetilde{\mathrm{HHI}}$ denote confidence advantage and concentration advantage.}
\label{fig:planner-behavior-combined}
\end{figure}

\noindent\textbf{Planning Stability.}
We examine whether the Planner's allocation is stable across repeated calls. For each parent order, we compute the KL dispersion of the quantity preference scores $a_n$ defined in Sec.~\ref{sec:long_horizon_planner} across eight repeated runs. Fig.~\ref{fig:planner-behavior-combined}(a) shows consistently modest KL dispersion for all three model settings, indicating broadly stable long-horizon score distributions across calls for the same parent order.

\noindent\textbf{Confidence-Based Allocation.}
We examine whether higher LLM confidence is associated with higher allocation concentration. We use the LLM confidence score $c$ and the Herfindahl--Hirschman Index ($\mathrm{HHI}$) of $a_n$ to measure confidence and allocation concentration, respectively. Following the eight repeated runs in experiments setup, for each parent order, we convert $c$ and $\mathrm{HHI}$ into advantages, $\tilde{c}$ and $\widetilde{\mathrm{HHI}}$, by subtracting their eight-run averages. Fig.~\ref{fig:planner-behavior-combined}(b) shows positive slopes for all three models, indicating that, for the same parent order, the LLM makes more concentrated allocations when it is more confident. This mirrors human investors: stronger self-belief leads to more concentrated portfolios \citep{barth2018costs}. The smaller slope for GPT-5.4 none suggests that the strength of this relation differs across models.

\noindent\textbf{Confidence and Performance.}
We examine whether LLM confidence predicts strategy performance. We regress $bp$ on confidence $c$, both with and without controls for parent-order direction, log quantity, execution-window length, and stock volatility; further details are provided in the supplementary material. Tab.~\ref{tab:confidence_performance_regression} shows significantly positive coefficients on $c$ across all regressions, indicating that higher LLM confidence is associated with better performance. This differs from human investors, whose overconfidence often leads to worse returns \citep{odean1999investors}. The larger coefficients for GPT-5.4 none suggest that this relation varies across models.

\subsection{Executor Behavior}

\noindent\textbf{Response to Time Pressure.}
We examine whether the Executor responds systematically to time pressure. We regress the LLM-generated quantity adjustment score $z$, defined in Eq.~\ref{eq:QE}, on time pressure $TP$, which measures how close the current time is to the end of the sub-plan. A larger $z$ means relatively more quantity allocated at the current decision time, and a higher $TP$ means the current sub-plan is closer to its deadline. Details are in the supplementary material.

Tab.~\ref{tab:executor_behavior_regression} shows that $TP$ has significantly negative coefficients across all model settings. This indicates that the Executor allocates more quantity when time pressure is still low, so it reduces the risk of having too much quantity left near the deadline. This differs from human behavior: humans often delay work until the deadline gets close and feel more urgent near the end \citep{steel2006integrating}.

\begin{table}[t]
\centering
\small
\setlength{\tabcolsep}{1.5pt}
\renewcommand{\arraystretch}{1.28}
\begin{tabular}{c*{6}{c}}
\toprule
& \multicolumn{2}{c}{GPT-5.4 none}
& \multicolumn{2}{c}{DS-v4-f none}
& \multicolumn{2}{c}{DS-v4-f high} \\
\midrule
$TP$
& -0.16$^{***}$
& -0.16$^{***}$
& -0.11$^{***}$
& -0.10$^{***}$
& -0.11$^{***}$
& -0.09$^{***}$ \\
{}
& (-11.42)
& (-11.27)
& (-5.13)
& (-4.86)
& (-6.77)
& (-6.41) \\
$LR$
& 10.49$^{***}$
& 10.69$^{***}$
& 1.72
& 2.26
& 3.26
& 3.61 \\
{}
& (4.86)
& (5.06)
& (0.53)
& (0.79)
& (1.13)
& (1.40) \\
Controls
& NO & YES
& NO & YES
& NO & YES \\
\midrule
Adj. $R^2$
& 0.019 & 0.034
& 0.013 & 0.082
& 0.018 & 0.097 \\
Obs.
& 44989 & 44989
& 44072 & 44072
& 43967 & 43967 \\
\bottomrule
\end{tabular}
\caption{Executor behavior regression. The dependent variable is $z$. $TP$ denotes time pressure, and $LR$ denotes the side-adjusted log return over the previous $\tau$ minutes. Parentheses report $t$-statistics based on parent-order-clustered standard errors.  $^{***}p<0.01$, $^{**}p<0.05$, and $^{*}p<0.1$.}
\label{tab:executor_behavior_regression}
\end{table}

\noindent\textbf{Beyond Heuristics.}
We further test whether LLM execution decisions reflect broader market reasoning or merely follow a simple recent market movement heuristic. We add $LR$ to the Executor regression. $LR$ is the log return over the previous $\tau$ minutes, with its sign reversed for sell orders, as detailed in the supplementary material. Tab.~\ref{tab:executor_behavior_regression} shows that $LR$ is significant positive for GPT-5.4 but insignificant for DS-v4-f. This suggests that GPT-5.4's execution decisions partly follow recent trends, whereas those of DS-v4-f cannot be explained by simple trend-following or mean-reversion. Since w/o Planner still outperforms TWAP in Tab.~\ref{tab:module_ablation}, these non-heuristic decisions by DS-v4-f provide genuine performance gains.

\section{Conclusion}

This paper presents the first study of LLMs for parent-order execution. We propose PACE, which separates long-horizon planning from short-horizon execution without pre-specified market assumptions or task-specific training. Experiments demonstrate consistent improvements over all baselines under both order-submission settings. Behavioral analysis further reveals that LLM decisions differ markedly from those of human investors. Higher model confidence predicts better performance, and the model trades earlier rather than deferring execution until the deadline. These patterns suggest that LLMs may help human traders make more disciplined execution decisions. Future work may evaluate PACE in live trading and enrich its inputs by incorporating news, market microstructure information, and cross-asset signals.

\bibliography{custom}

\section{Appendix}

\subsection{Notation}
\label{appendix:notion}

The notation used in this paper is summarized in Table~\ref{tab:notation}.

\begin{table*}[t]
\centering
\small
\setlength{\tabcolsep}{4.5pt}
\renewcommand{\arraystretch}{1.05}
\begin{tabular}{p{0.075\textwidth}p{0.385\textwidth}p{0.075\textwidth}p{0.385\textwidth}}
\toprule
Symbol & Meaning & Symbol & Meaning \\
\midrule
$O$ & Parent order & $s$ & Stock ID \\
$d$ & Trading direction & $Q$ & Total execution quantity \\
$t_s$ & Parent-order start time & $t_e$ & Parent-order end time \\
$T$ & Parent-order execution-window duration & $\Delta$ & Fixed decision interval \\
$\tau$ & Fixed sub-plan duration & $K$ & Number of decision times \\
$t_k$ & The $k$-th decision time & $L$ & Lookback-window length \\
$\mathcal{H}_{t-L:t}$ & Market history over $[t-L,t]$ & $P_i$ & Mid-price at minute $i$ \\
$V_i$ & Total traded volume at minute $i$ & $A_i$ & Ask1 price at minute $i$ \\
$B_i$ & Bid1 price at minute $i$ & $Q^{\mathrm{TWAP}}$ & TWAP quantity at each decision time \\
$\mathcal{C}^{\mathrm{TWAP}}$ & TWAP curve & $N$ & Number of Planner time slots \\
$S_n$ & The $n$-th execution sub-plan & $u_n$ & Start time of sub-plan $n$ \\
$v_n$ & End time of sub-plan $n$ & $q_n$ & Execution quantity of sub-plan $n$ \\
$\tau_n$ & Duration of sub-plan $n$ & $\mathcal{R}^{L}$ & Planner's long-horizon trend assessment \\
$a_n$ & Quantity preference score for slot $n$ & $c$ & Planner confidence score \\
$w_n$ & Planner allocation weight for slot $n$ & $\lambda$ & LLM-allocation strength coefficient \\
$z_t$ & Executor quantity adjustment score at time $t$ & $Q^E_t$ & Executor order quantity at time $t$ \\
$\gamma$ & Executor deviation coefficient from TWAP & $\mathcal{D}$ & Market snapshots \\
$\pi$ & Execution strategy & $P^{\mathrm{TWAP}}$ & TWAP price benchmark \\
$P^{\mathrm{Trade}}$ & Average execution price & $Q^{\mathrm{Trade}}$ & Traded parent-order quantity \\
$cr$ & Parent-order completion rate & $bp$ & Basis-point price performance \\
$wbp$ & Value-weighted basis-point performance & $TP$ & Time pressure \\
$LR$ & Side-adjusted recent log return & $\mathrm{HHI}$ & Allocation concentration \\
$\tilde{c}$ & Confidence advantage & $\widetilde{\mathrm{HHI}}$ & Concentration advantage \\
$M$ & Number of parent orders & $R$ & Number of repeated runs \\
$\kappa$ & AC schedule front-loading parameter & $\eta$ & ML baseline adjustment coefficient \\
\bottomrule
\end{tabular}
\caption{Notation used in this paper.}
\label{tab:notation}
\end{table*}

\subsection{Parent Order Generation}
\label{sec:pogeneration}

For each trading day, we randomly generate 10 parent orders to evaluate different execution strategies under the same parent orders. The start time of each parent order is fixed at 10:30:00, and the end time is determined by a randomly sampled execution window. The execution window is sampled from \{10, 20, 30, 40, 50, 60\} minutes, so the end time ranges from 10:40:00 to 11:30:00. The order side is randomly sampled from \{BUY, SELL\}. The parent-order quantity is randomly generated between 100 and 10,000 shares and is constrained to be a multiple of 100 shares. Among the generated parent orders, 51\% are buy orders and 49\% are sell orders. Other summary statistics are shown in Table~\ref{tab:parent_order_stats}.

\begin{table}[t]
\centering
\small
\begin{tabular}{lcc}
\toprule
Statistic & $Q$ (shares) & $T$ (minutes) \\
\midrule
Mean & 5000.48 & 35.29 \\
Std. & 2878.38 & 16.40 \\
Min & 100 & 10 \\
25\% & 2500 & 20 \\
50\% & 4900 & 40 \\
75\% & 7500 & 50 \\
Max & 10000 & 60 \\
\bottomrule
\end{tabular}
\caption{Summary statistics of generated parent orders.}
\label{tab:parent_order_stats}
\end{table}

\subsection{Robustness Analysis}
\label{sec:additional_robustness}

\noindent\textbf{Statistical robustness.}
We use 5,000 bootstrap resamples over parent orders to test whether PACE's gains are statistically robust. Table~\ref{tab:bootstrap_robustness} shows that DS-v4-f has positive 95\% confidence intervals under both order-submission settings, while GPT-5.4 remains close to the significance boundary. These results support the statistical robustness of the strongest PACE variant.

\begin{table}[t]
\centering
\small
\setlength{\tabcolsep}{4.5pt}
\renewcommand{\arraystretch}{1.12}
\begin{tabular}{llrrr}
\toprule
Setting & Model & gain & 95\% CI & $p$-value \\
\midrule
Aggressive & GPT-5.4 & +0.52 & [0.00, 1.05] & 0.03 \\
Aggressive & DS-v4-f & +1.02 & [0.15, 2.12] & 0.00 \\
Passive & GPT-5.4 & +0.50 & [0.00, 1.00] & 0.03 \\
Passive & DS-v4-f & +1.07 & [0.05, 2.24] & 0.01 \\
\bottomrule
\end{tabular}
\caption{Bootstrap robustness test for PACE. Gains are computed relative to TWAP.}
\label{tab:bootstrap_robustness}
\end{table}

\noindent\textbf{Depth-limited matching.}
We test whether PACE remains effective under a stricter matching rule. In this setting, an aggressive order can trade only up to the available Ask1 or Bid1 volume, and the remaining quantity is cancelled. Table~\ref{tab:depth_limited_matching} shows that PACE still outperforms TWAP under this more conservative matching mechanism.

\begin{table}[t]
\centering
\small
\setlength{\tabcolsep}{7pt}
\begin{tabular}{ccc}
\toprule
Strategy & Original matching & Depth-limited matching \\
\midrule
TWAP & -3.28 & -3.08 \\
PACE & \textbf{-2.26} & \textbf{-2.41} \\
\bottomrule
\end{tabular}
\caption{Depth-limited matching test using DS-v4-f. Values are reported in $wbp$.}
\label{tab:depth_limited_matching}
\end{table}

\noindent\textbf{Final-sweep analysis.}
We examine whether passive-setting gains are driven by the aggressive sweep in the final minute. Table~\ref{tab:passive_sweep_analysis} shows that DS-v4-f achieves -3.64 $wbp$ before the sweep, better than its overall performance of -3.92 $wbp$, while its sweep trades achieve -6.63 $wbp$. Thus, the final sweep reduces rather than creates its performance gain.

\begin{table}[t]
\centering
\small
\setlength{\tabcolsep}{5pt}
\begin{tabular}{lrrr}
\toprule
Strategy & Sweep (\%) & Pre-sweep & Sweep \\
\midrule
TWAP    & 14.93 & -4.80 & -7.38 \\
AC      & 14.60 & -4.55 & -7.15 \\
XGB     & 13.64 & -4.44 & -7.68 \\
LSTM    & 13.62 & -4.39 & -10.50 \\
GPT-5.4 & 18.05 & -4.21 & -7.47 \\
DS-v4-f & 18.80 & \textbf{-3.64} & -6.63 \\
\bottomrule
\end{tabular}
\caption{Final-sweep analysis under passive submission. The last two columns report $wbp$; Sweep (\%) is the fraction of executed quantity traded by the final-minute sweep.}
\label{tab:passive_sweep_analysis}
\end{table}

\noindent\textbf{Repeated-run stability.}
Table~\ref{tab:run_stability} reports $wbp$ over eight repeated runs for each parent order. The small across-run variation indicates that PACE is stable across repeated LLM calls.

\begin{table}[t]
\centering
\small
\setlength{\tabcolsep}{4.8pt}
\renewcommand{\arraystretch}{1.12}
\begin{tabular}{lcccc}
\hline
\multirow{2}{*}{Statistic}
& \multicolumn{2}{c}{Aggressive}
& \multicolumn{2}{c}{Passive} \\
\cline{2-5}
& GPT-5.4 & DS-v4-f & GPT-5.4 & DS-v4-f \\
\hline
Mean   & -2.76 & -2.26 & -4.49 & -3.92 \\
Std.   &  0.24 &  0.14 &  0.34 &  0.21 \\
Min.   & -3.32 & -2.45 & -5.08 & -4.24 \\
Median & -2.72 & -2.22 & -4.42 & -3.84 \\
Max.   & -2.48 & -2.04 & -4.06 & -3.58 \\
\hline
\end{tabular}
\caption{
Repeated-run stability of PACE.
Each statistic is computed over eight repeated runs, and values are reported in $wbp$.
}
\label{tab:run_stability}
\end{table}

\subsection{Baseline Details}
\label{sec:baseline}

\subsubsection{AC Strategy}

We implement an Almgren-Chriss strategy as a traditional execution baseline. Specifically, given the total parent-order quantity $Q$ and execution-window length $T$, the AC strategy first computes the target remaining inventory at time $t$ as follows:

\begin{equation}
x_{t}=Q \frac{\sinh(\kappa(T-t))}{\sinh(\kappa T)}
\end{equation}

where $x_{t}$ denotes the remaining quantity to be executed after time $t$, and $\kappa$ controls the degree of front-loading in the execution curve. When $\kappa=0$, the strategy reduces to an approximately uniform TWAP execution curve. As $\kappa$ increases, the strategy completes more trading quantity earlier in the execution window, leading to a more front-loaded execution pattern. Based on the target remaining inventory, the strategy further obtains the target cumulative executed quantity at each minute:

\begin{equation}
Q^{AC}_t=Q-x_{t}
\end{equation}

In the experiments, we perform a grid search over $\kappa \in \{0.1,0.3,0.5,0.8,1.0\}$ and use the best-performing parameter $\kappa=0.5$.

\subsubsection{Learning-based Strategies}
\label{sec:mlstrategy}

To further compare PACE with learning-based strategies, we construct two machine learning baselines: XGB and LSTM. Both models are trained to predict the next-minute price direction, and the prediction signal is used to adjust the TWAP quantity at each execution step.

The models are trained on market-wide minute-level data from January 1, 2026 to March 31, 2026 and evaluated on April 2026 data used in the main experiments, ensuring that the training period precedes the evaluation period. The input features include log return, log-return volatility, order imbalance, trade imbalance, and spread volatility. The prediction target is a three-class direction label of the next-minute return: up, down, or flat.

During backtesting, both baselines use TWAP as the default execution curve. For a buy parent order, the strategy increases the current TWAP quantity when the model predicts an upward price movement; for a sell parent order, it increases the current TWAP quantity when the model predicts a downward price movement.

\begin{equation}
Q^{\mathrm{ML}}_t =
\begin{cases}
(1+\eta)Q^{\mathrm{TWAP}}_t, & y_t=1,\\
Q^{\mathrm{TWAP}}_t, & y_t=0,
\end{cases}
\label{eq:ml_adjusted_twap}
\end{equation}

where $y_t=1$ indicates that the current signal is favorable, and $y_t=0$ denotes all other cases. The adjustment magnitude is controlled by the hyperparameter $\eta$. We search over $\eta \in \{0.1,0.3,0.5,0.8,1.0\}$ and report the best result in the main experiments.

\subsection{Behavioral Analysis Details}

\subsubsection{Planner Stability}
\label{sec:planner_behavior1}

We evaluate whether the Planner gives stable long-horizon quantity preference scores when the same parent order is queried repeatedly. For parent order $i$ and repeated run $r$, the Planner outputs slot scores
$\mathbf{a}_{i,r}=(a_{i,r,1},\ldots,a_{i,r,N_i})$.
We convert these scores into a score-implied allocation distribution using a fixed softmax transformation:

\begin{equation}
p_{i,r,n}
=
\frac{\exp(a_{i,r,n})}
{\sum_{m=1}^{N_i}\exp(a_{i,r,m})}
\end{equation}

For each parent order, we compute the average score distribution across its $R=8$ repeated runs:

\begin{equation}
\bar{p}_{i,n}
=
\frac{1}{R}
\sum_{r=1}^{R} p_{i,r,n}
\end{equation}

The KL dispersion of run $r$ is then defined as the divergence between its score distribution and the parent-order average distribution:

\begin{equation}
D_{i,r}
=
D_{\mathrm{KL}}(p_{i,r}\Vert \bar{p}_{i})
=
\sum_{n=1}^{N_i}
p_{i,r,n}
\log
\frac{p_{i,r,n}}{\bar{p}_{i,n}}
\end{equation}

The parent-order-level KL dispersion is the average across repeated runs:

\begin{equation}
\mathrm{KL}_{i}
=
\frac{1}{R}
\sum_{r=1}^{R}
D_{i,r}
\end{equation}

A lower value means that the Planner produces more similar long-horizon score distributions for the same parent order across repeated runs. The model-level statistic reported in Fig.7(a) of the main paper is the average of $\mathrm{KLDispersion}_{i}$ across parent orders.

\subsubsection{Confidence and Concentration}
\label{sec:planner_behavior2}

We examine whether higher Planner confidence is associated with more concentrated long-horizon allocation preferences. We use the same score-implied allocation distribution $p_{i,r,n}$ defined in Appendix~\ref{sec:planner_behavior1}. For each parent order $i$ and repeated run $r$, score concentration is measured by the HHI:

\begin{equation}
\mathrm{HHI}_{i,r}
=
\sum_{n=1}^{N_i}
p_{i,r,n}^{2}
\end{equation}

A larger HHI means that the Planner's long-horizon preference is more concentrated on fewer slots. To compare repeated runs of the same parent order, we remove the parent-order mean from both confidence and HHI:

\begin{equation}
\tilde{c}_{i,r}
=
c_{i,r}
-
\frac{1}{R}
\sum_{r'=1}^{R}
c_{i,r'}
\end{equation}

\begin{equation}
\widetilde{\mathrm{HHI}}_{i,r}
=
\mathrm{HHI}_{i,r}
-
\frac{1}{R}
\sum_{r'=1}^{R}
\mathrm{HHI}_{i,r'}
\end{equation}

The demeaning step controls for parent-order-level differences. Therefore, Fig.7(b) of the main paper compares whether a run that is more confident than the same parent order's average also has a more concentrated score-implied allocation distribution.

\subsubsection{Confidence and Performance}
\label{sec:planner_behavior3}

We examine whether Planner confidence predicts execution performance. The main explanatory variable is the Planner confidence score $c$. We use $bp$ as dependent variable. 

We estimate the following OLS specification:

\begin{equation}
bp
=
\alpha
+
\beta c
+
\mathbf{X}'\Gamma
+
\epsilon
\end{equation}

where the control vector $\mathbf{X}$ includes parent-order side, log parent-order quantity, execution-window length, and stock volatility. We report both specifications without controls and specifications with controls in Table.4 of the main paper.

\subsubsection{Executor Regression}
\label{appendix:exe_reg}

We examine whether the Executor's quantity adjustment responds to time pressure and recent price movements. The dependent variable is the LLM-generated quantity adjustment score $z$. A larger $z$ means that the Executor allocates more quantity.

For each Executor decision, we define time pressure as:

\begin{equation}
TP_t
=
1-\frac{\tau^{remaining}_t}{\tau}
\end{equation}

where $\tau^{remaining}_t$ is the remaining time of sub-plan at time $t$, and $\tau$ is the total duration of sub-plan. A larger $TP_t$ means that the current time is closer to the sub-plan deadline.

We also define $LR$ as the side-adjusted log return over the previous $\tau$ minutes:

\begin{equation}
LR_t =
\begin{cases}
\log P_t - \log P_{t-\tau}, & d=\mathrm{BUY},\\
-\left(\log P_t - \log P_{t-\tau}\right), & d=\mathrm{SELL},
\end{cases}
\end{equation}

where $P_t$ is the mid-price at time $t$, and $d$ is the parent-order direction.

We estimate the following OLS specification:

\begin{equation}
z_t
=
\alpha
+
\beta_1 TP_t
+
\beta_2 LR_t
+
\mathbf{X}'\Gamma
+
\epsilon
\end{equation}

The control vector $\mathbf{X}$ includes parent-order side, log parent-order quantity, execution-window length, stock volatility, and log sub-plan quantity. 
Tab.5 reports both specifications without controls and specifications with controls.

\subsection{Prompts}
\label{sec:prompt}

The prompt details are shown in Figure~\ref{fig:planner_prompt} and Figure~\ref{fig:executor_prompt}.

\begin{figure*}[t]
\centering
\begin{tcolorbox}[
  width=0.98\textwidth,
  colback=gray!4,
  colframe=gray!55,
  boxrule=0.4pt,
  arc=1mm,
  left=2mm,
  right=2mm,
  top=1.5mm,
  bottom=1.5mm
]
\small
\textbf{Planner Prompt Template}

\medskip

You are a professional trading planner. Based on the following market information and rules, score the fixed future slots so the system can convert your scores into an execution plan.

\medskip

\textbf{Input:}
minute-level historical mid-price and traded volume; parent-order direction, total quantity, and duration; fixed 5-minute future slots.

\medskip

\textbf{Rules:}
The execution window is split into fixed 5-minute slots. The model must score every slot instead of outputting raw quantities or custom time ranges. Valid scores are $-1,-0.5,0,0.5,1$, where a higher score means that the slot deserves more allocation. If the signal is weak or noisy, the model should use mostly zero scores and a low confidence value so the final allocation stays close to TWAP. The goal is to obtain an average execution price better than the TWAP price: lower for buy orders and higher for sell orders.

\medskip

\textbf{Response format:}
First provide brief analysis reasons, then output a JSON object with a short summary, a confidence value, and one score for each fixed slot.

\medskip

\textbf{Planner JSON:}
\[
\{\text{summary},\ \text{confidence},\ \text{slot\_scores}:[\{\text{slot\_index},\ \text{score}\}]\}
\]

\medskip

\textbf{Constraints:}
Output exactly one score for each fixed future slot. Do not output raw quantities or time ranges.
\end{tcolorbox}
\caption{Condensed Planner prompt template. Runtime variables such as market history, parent-order fields, and slot definitions are filled dynamically for each parent order.}
\label{fig:planner_prompt}
\end{figure*}

\begin{figure*}[t]
\centering
\begin{tcolorbox}[
  width=0.98\textwidth,
  colback=gray!4,
  colframe=gray!55,
  boxrule=0.4pt,
  arc=1mm,
  left=2mm,
  right=2mm,
  top=1.5mm,
  bottom=1.5mm
]
\small
\textbf{Executor Prompt Template}

\medskip

You are a financial execution assistant. Based on the recent market information, the Planner's long-term intent, and the current TWAP baseline quantity, determine how the current minute's order quantity should be adjusted.

\medskip

\textbf{Input:}
current time; parent-order start and end time; total parent-order quantity; minutes remaining; executed quantity; in-flight pending quantity; current TWAP baseline quantity; Planner's long-term prediction; recent minute-level mid-price and traded volume.

\medskip

\textbf{Rules:}
The model should judge whether the short-term market condition is favorable, unfavorable, or unclear relative to the current TWAP execution. For a buy order, if the price is expected to rise, trading more now is preferred; if the price is expected to fall, trading less now is preferred. For a sell order, the direction is reversed. If the signal is weak, noisy, or conflicts with the Planner's long-term intent, the model should choose a neutral adjustment.

\medskip

\textbf{Response format:}
First provide brief analysis reasons, then output a JSON object containing only the adjustment level.

\medskip

\textbf{Executor JSON:}
\[
\{\text{adjustment\_level}: z_t\}
\]

\medskip

\textbf{Valid adjustment levels:}
$z_t \in \{-1,-0.5,0,0.5,1\}$, where $0$ means following the TWAP baseline quantity, positive values mean trading more than TWAP, and negative values mean trading less than TWAP.

\medskip

\textbf{Constraints:}
Do not output raw quantities while respecting remaining quantity, pending orders, and the parent-order completion constraint.
\end{tcolorbox}
\caption{Condensed Executor prompt template. Runtime variables such as recent market history, Planner intent, TWAP baseline quantity, and execution progress are filled dynamically at each decision minute.}
\label{fig:executor_prompt}
\end{figure*}

\subsection{AI Assistance Statement}

We used AI assistants as supporting tools for code development, document editing, and language polishing. All research questions, experimental designs, implementation decisions, analytical conclusions, and final manuscript statements were reviewed, verified, and are the responsibility of the authors.

% \appendix
% \input{secs/08_appendix}

\end{document}